\documentclass[letterpaper,conference]{IEEEtran}
\usepackage{lipsum,graphicx,multicol}
\usepackage{algorithm, algorithmic}

\newtheorem{theorem}{Theorem}

\newtheorem{lemma}{Lemma}

\usepackage{ifpdf}

\ifCLASSINFOpdf
\else
\fi
\usepackage{url}
\usepackage{amsmath,graphicx}
\usepackage{placeins}
\usepackage{amsmath,epsfig}
\usepackage{epstopdf}
\usepackage{amssymb}

\usepackage{dblfloatfix}
\usepackage{amssymb}
\usepackage{caption}
\usepackage{subfigure}
\usepackage{pstricks}
\usepackage{subfloat}
\usepackage{lipsum}

\begin{document}
%
\title{QoE-Aware Resource Allocation for Small Cells}
\author{\IEEEauthorblockN{}
\IEEEauthorblockA{Anis Elgabli, Ali Elghariani*, Vaneet Aggarwal, Mark Bell\\
Purdue University, West Lafayette IN 47907\\ 
*University of Tripoli }}

\IEEEtitleabstractindextext{%
\begin{abstract}
In this paper, we study the problem of Quality of Experience (QoE) aware resource allocation in wireless systems. In particular, we consider application-aware joint Bandwidth-Power allocation for a small cell. We optimize a QoE metric for multi-user video streaming in a small cell that maintains a trade-off between maximizing the playback rate of each user and ensuring proportional fairness (PF) among users. We formulate the application-driven joint bandwidth-power allocation as a non-convex optimization problem. However, we develop a polynomial complexity algorithm, and we show that the proposed algorithm achieves the optimal solution of the proposed optimization problem. Simulation results show that the proposed QoE-aware algorithm significantly improves the average QoE. Moreover, it outperforms the weighted sum rate allocation which is the state-of-the-art physical resource allocation scheme.\\

\end{abstract}

\begin{IEEEkeywords}
QoE, Video streaming, Bandwidth-Power allocation, non-convex optimization
\end{IEEEkeywords}}

\maketitle
\IEEEdisplaynontitleabstractindextext

\section{Introduction}
\label{sec:intro}
Video streaming is the dominant contributor to the cellular traffic. Currently, video streaming accounts for $50\%$ of cellular traffic and it is expected to to grow to $75\%$ of the mobile data traffic by the year of 2020 \cite{ericsson_report}. This increase has forced service providers to enhance their infrastructures to support high-quality video streaming. Despite these efforts, users frequently experience low Quality-of-Experience (QoE) metrics such as choppy videos and playback stalls~\cite{zou2015can}.

In the modern video coding, the video is divided into chunks with $L$ second duration each. Each chunk is either encoded into $N$ {\em independent} versions with different qualities, or encoded into ordered \emph{layers}, i.e, one \emph{base layer} (Layer 0) with the lowest playable quality, and multiple \emph{enhancement layers} (Layer $i>$0) that further improve the chunk quality. As an example of the former technique is H.264/MPEG-4 AVC (Advanced Video Coding) which was standardized in 2003~\cite{H264}. Moreover, an example that represents the layered encoding technique is Scalable Video Coding (SVC) which was standardized in 2007 as an extension to H.264~\cite{4317636}. In both encoding techniques, the video is available at different rates, and a rate adaptation logic needs to make a decision about the quality of the next few chunks to request based on the current buffer and/or bandwidth prediction. 

The recent adoption of the open standard MPEG-DASH ~\cite{DASH} has made Adaptive Bit Rate (ABR) streaming the most popular video streaming solution. Commercial systems such as Apple's HLS~\cite{HLS}, Microsoft's Smooth Streaming~\cite{SS}, and Adobe's HDS~\cite{HDS} are all ABR streaming algorithms. In recent studies, researchers have investigated various approaches for making streaming decisions, for example, by using control theory~\cite{MPC,Miller15}, Markov Decision Process~\cite{Jarnikov11}, machine learning~\cite{Claeys14}, client buffer information~\cite{BBA}, and data-driven techniques~\cite{Liu12,C3,CS2P}. However, all of these rate adaptation techniques resides at the application layer of the client. Hence, the client decides based on the bandwidth prediction and/or the current buffer size what rate should next chunk of the video be fetched at. Therefore, the wireless base station provides no help in ensuring a certain rate to the client.  In this paper, we still assume that each user is running his/her own rate adaptation logic at the application layer to adapt to the network changes. The QoE aware resource allocation algorithm proposed in this paper does not replace the application layer rate adaptation technique that runs at the client side. However, it allows the base station (BS) to consider the application requirements in assigning the resources. Therefore, the physical layer resources (bandwidth and power) can be allocated such that the average QoE among users is maximized.

In this paper, driven by the following aspects; (i) most of the cellular traffic is video, (ii) the video is available at a specific set of rates which is a discrete in nature, and (iii) the new paradigm of a small cell in which a small BS with low power and more customizability resides in a small place can serve few users, we propose a video QoE-Aware bandwidth-power allocation algorithm for a small cell scenario.


To motivate our problem, consider a scenario in which $U$ users are connected to a small cell and watching videos. The videos are encoded into $N+1$ rates; $r_0, r_1,...,r_N$. If a user cannot achieve a rate of $r_0$, then the user will run into re-buffering which significantly degrades the user's QoE. The user's QoE is generally evaluated by the mean opinion score (MOS) \cite{singh2012interference,rec1996p}, which practically ranges, e.g., from 1 to 4.5. Moreover, video MOS is shown to be bound logarithmic with respect to the Quality of Service (QoS) parameter \cite{saul2009multiuser,li2012qoe}. The QoS parameter in video streaming is the video playback rate. Therefore, the increase in the QoE is more significant when switching from rate $r_{n-1}$ to $r_{n}$ compared to switching from $r_{n}$ to $r_{n+1}$. Hence, in the case of multiple users connected to a small cell scenario, pushing a large number of users to the rate $r_n$ is more preferable than just increasing the rates of only a few users to rates higher than $ r_n$ at the cost of dropping the rates of the majority of users to rates below $ r_n$. This strategy improves the playback rate of each user and ensures proportional fairness among users.

In the physical layer, utility-based resource allocation in wireless networks that ensures fairness among users with different application requirements was studied in \cite{kuo2007utility}. However, the authors in \cite{kuo2007utility} assumes that the traffic of the users with hard QoS are available at a single rate which is not true for the modern videos. Moreover, all users are sharing a single total rate which does not hold when the users are experiencing different channels. The authors of \cite{cho2015qoe} proposed a QoE-based proportional fairness utility function in which a continuously differentiable
MOS function is maximized based on relaxing the discrete constraints to simplify the problem. In contrast to the previous work, we exploit the discrete nature of the available video rates and show that the non-convex optimization problem is solvable in polynomial time.

In this paper, we first formulate the joint bandwidth-power allocation for a video traffic in a multi-user small cell scenario as a non-convex optimization problem. The  objective is to optimize a QoE metric that maintains a trade-off between maximizing the playback rate of each user and ensuring proportional fairness among all users. Secondly, we develop a novel algorithm that solves this problem optimally in a polynomial time complexity. Finally, we compare our proposed algorithm with the weighted sum rate, which is the state-of-the-art physical resource allocation technique. Moreover, we show that our algorithm outperforms the weighted sum rate in terms of avoiding rebuffering times and  maintaining high QoE for more users. 

\section{System Model}
\begin{figure}
 \centering
\includegraphics [width=6cm, height=4cm]{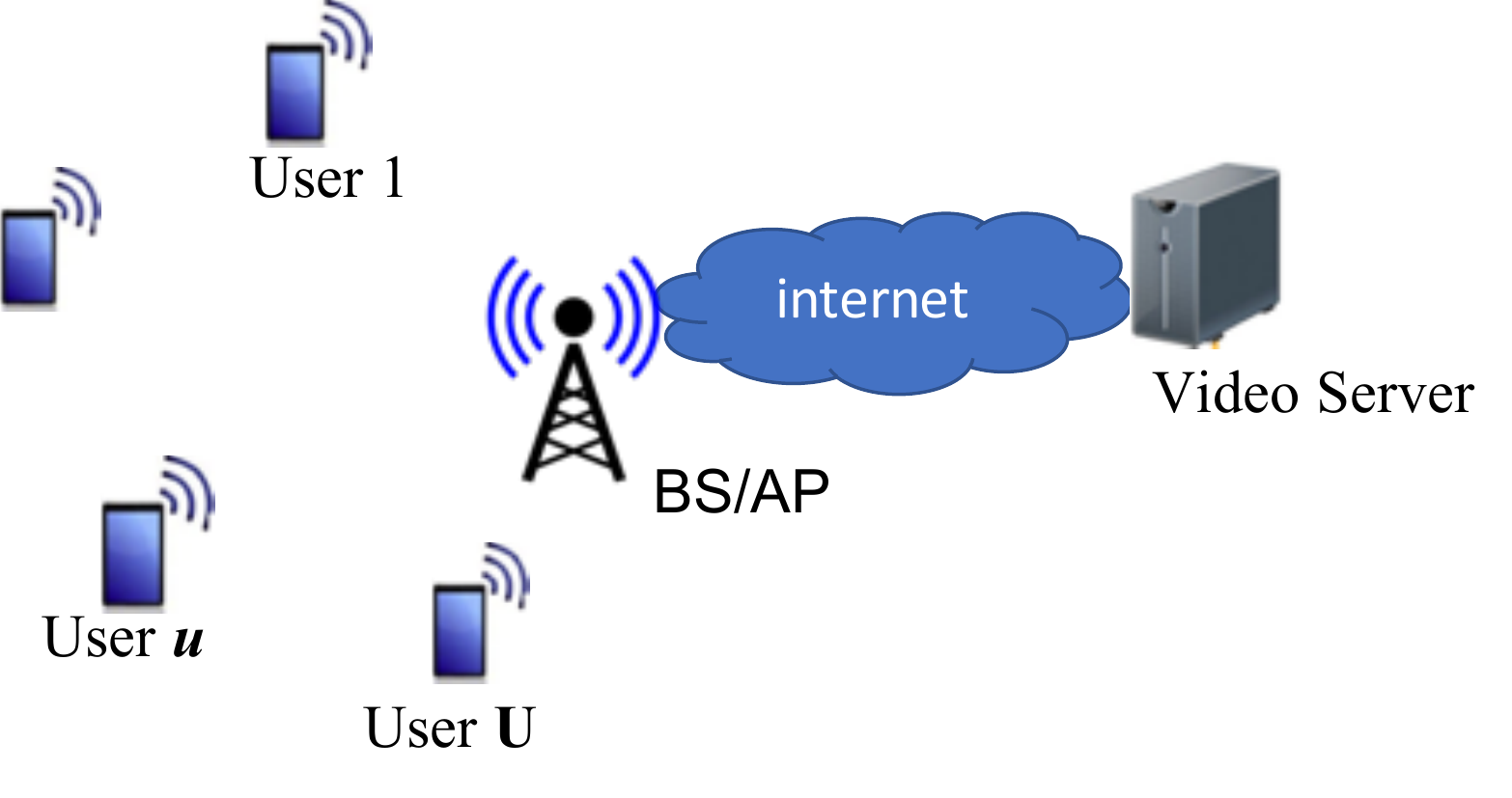}
\caption{ A small cell with multi users streaming in the downlink }
\label{fig:1}
\end{figure}

Consider a single cell wireless network as shown in Fig. \ref{fig:1}, with $U$ users,  $ {u} \in \mathcal{U} =\{1,2,3, \dots, U\} $. A spectrum of a total bandwidth, $B$, is available for the downlink transmissions from the Base Station (BS) or Access Point (AP) to all users. Each user is allocated a downlink portion of the spectrum for online video steaming. This spectrum is assumed to be flat fading  and can be divided into distinct and non-overlapping channels of unequal bandwidths, so that the users share the available spectrum through a frequency division manner. Let $p_u$ and $b_u$ represent the allocated transmit power and channel bandwidth of the BS to serve the $u$-th user. The received SNR at the $u$-th user is $\text{SNR}_u=\frac{ p_u H_{u}} {N_{o} b_u}$,  and according to the Shannon theorem, the channel capacity, $C_u$, of the $u$-th user is: $ C_{u}=b_{u} log ( 1+ \frac{ p_{u} {H_u}}{N_{o} b_{u}})$, where  $N_o$ stands for the power spectral density of additive white Gaussian noise (AWGN), and $H_u$ denotes the channel power of the link between the BS and the $u$-th user, which can be modeled based on both large-scale and small-scale fading effects as follows: \\
\begin{equation}
H_u = | \alpha_u h_{u}|^2
\label{eq: 1}
\end{equation} 
where $h_{u}$ represents the small scale fading  which is modeled as $ \mathbb{C}\mathcal{N} (0, 1) $. $\alpha_u $ represents the large scale propagation effect for the $u$-th user with $ \alpha_u = \sqrt( c/{d_u^m})$. $d_u$ is the relative distance between the $u$-th user and the base station, $m$ is the path loss exponent which is typically between 1.6 and 4 depending on the environment, and $c$ is a constant related to the propagation loss and antenna gains. Large scale fading that results from shadowing can be modeled as a log normal distribution for the $u$-th user channel, but since in this paper we focus on a small cell such as the one located inside a building, the shadowing effect can be ignored. 
In this model, we assume that each user watches a video which is divided into $C$ chunks with $L$-second duration each. This video is encoded at one of the $N+1$ playback rates, $r_0,\cdots r_{N}$. Higher playback rate leads to a better QoE. However, higher playback rate requires more physical resources (Bandwidth and Power). We denote ${\cal B}\triangleq \{b_u\;|\; b_u\triangleq [b_1,b_2,\ldots,b_U]\}$ as the {\em bandwidth allocation}, ${\cal P}\triangleq \{p_u\;|\; p_u\triangleq [p_1,p_2,\ldots,p_U]\}$ as the {\em Power allocation}

\section{Problem Formulation}
In this section, we describe our problem formulation. Let $I_n^u$ be the decision variable of the $n$-th rate and $u$-th user, i.e, 
\begin{equation}\label{equ:fetchingPolicy}
\left\{\begin{array}{l}
I_n^u=1, \text{ if the $u$-th user is a candidate to the $n$-th rate}\\
I_n^u=0,  \text{ otherwise}\\
\end{array}\right.
\end{equation}
\begin{equation}
n \in\{0....N\} \nonumber
\end{equation}
Our objective function is a weighted sum of the decision variables. Let $\lambda_n$ be the weight of the $n$-th rate. In order to prefer pushing all the users to the $n$-th rate over any other choice that is pushing some users to the higher rates with the cost of dropping the rate of others to below the $n$-th rate, the weights should satisfy the following constraint.
\begin{equation}
\lambda_n > U \cdot \sum_{k=n+1}^N \lambda_k 
\label{lambdaCons}
\end{equation}
Equation \eqref{lambdaCons} states that the utility that is achieved by pushing a user to the $n$-th rate if possible is higher than the utility that is achieved by pushing all other users to rates $> n$-th rate. Therefore, maximizing this objective will maximize the playback rate of each user subjected to ensuring proportional fairness among users. The overall optimization problem is formulated as follows:
 \begin{equation}
 \textbf{Maximize} \sum_{n=0}^N\lambda_n\sum_{u=1}^U I_n^u 
 \label{equ:mainObj}
 \end{equation}
 subject to
  
 \begin{equation}
  I_0^u={\bf 1}\Big(b_ulog(1+\frac{\gamma p_u|\alpha_u h_u|_2^2}{N_ob_u}) \geq r_0\Big), \forall u
  \label{equ:eq1c1}
   \end{equation}

   \begin{equation}
   I_n^u=I_{n-1}^u \cdot {\bf 1}\Big(b_ulog(1+\frac{\gamma p_u|\alpha_u h_u|_2^2}{N_ob_u}) \geq r_n\Big), \forall u, n=1...N
   \label{equ:eq1c2}
   \end{equation}
   \begin{equation}
   \sum_u p_u \leq P_{max}
   \label{equ:eq1c3}
   \end{equation}
    \begin{equation}
   \sum_u b_u \leq B
   \label{equ:eq1c4}
   \end{equation}
   
    \begin{equation}
   I_n^u \in\{0,1\}, n=1...N,u
   \label{equ:eq1c5}
   \end{equation}
Where $\textbf{1}(x>y) =\{1, \text{if} \, x>y, \text{and} \, 0 \, \text{otherwise}\}$, $\gamma \in (0, 1]$ is the rate to the capacity gap, and $\lambda_{0,\cdots, N}$ must satisfy \eqref{lambdaCons}. Constraint \eqref{equ:eq1c1} states that $I_0^{(u)}$ is equal to 1 only if there is a bandwidth-power allocation such that the user $u$ achieves a rate $\geq r_0$. Constraint \eqref{equ:eq1c2} states that $I_n^{(u)}, \forall u, n > 0$ is equal to 1 only if the user $u$ is a candidate to the immediate lower rate ($I_{n-1}^{(u)}$) and there is a bandwidth-power allocation such that the user $u$ achieves a rate $\geq r_n$. \eqref{equ:eq1c3} and \eqref{equ:eq1c4} are the total power and bandwidth constraints. Finally, \eqref{equ:eq1c5} is the non-convexity constraint on the feasible set of $I_n^{(u)},\forall u,n$. Therefore, the optimization problem \eqref{equ:mainObj}-\eqref{equ:eq1c5} is a non-convex. In the next section, we will describe a novel algorithm that can solve this problem optimally in polynomial complexity. 

\section{Proposed Algorithm}
In this section, we explain the proposed algorithm which is listed in Table.~\ref{table:bbm}. We process rate levels in order such that when making the $n$-th rate level decisions, we only consider users who were candidates to the $(n-1)$-th rate level. i.e, we consider a user $u$ if $I_{n-1}^{u}=1$, otherwise, the constraint (\ref{equ:eq1c2}) will be violated. Let's index the users according to their channel gains where the user with the highest channel gain is given the index $1$, and the user with the lowest channel gain is given the index $U$. 

We start with rate $r_0$ by considering only user 1. Then we check if the user 1 is allocated the whole bandwidth and power ($b_1=B$, and $p_1=P_{max}$), would he/she be able  to achieve the $0$-th rate. i.e, If $b_1 log(1+\frac{\gamma p_1|h_1|_2^2}{N_ob_1}) \geq r_0$, then the user is a candidate to the $0$-th rate, and thus $I_0^{(1)}=1$. Otherwise, no user will be a candidate to this rate and the algorithm should terminate. In general if the $u$-th user is not a candidate to the $n$-th rate, every user $\in \{u+1,\dots, U\}$ will not be a candidate to rate levels $\geq n$-th rate. Hence the algorithm proceeds by checking the possibility of finding candidate users to the next rate among users 1 to $u-1$. If user 1 is a candidate to the $0$-th rate (i.e $I_0^{(1)}=1$), we solve the following optimization problem to find if user 2 can also be a candidate to the rate $r_0$ given that user 1 is already a candidate to this rate level. Note, the candidacy of user 1 to rate $r_0$ is formally stated by constraint~\eqref{equ:eq3c1}:

\begin{equation}
 \underset{b_1,b_2,p_1,p_2}{\textbf{argmax}} b_2log(1+\frac{\gamma p_2|\alpha_2 h_2|_2^2}{N_ob_2}) 
 \label{equ:Obj2}
 \end{equation}
 subject to
  \begin{equation}
 b_1log(1+\frac{\gamma p_1|\alpha_1 h_1|_2^2}{N_ob_1})\geq r_0
    \label{equ:eq3c1}
   \end{equation}
 \begin{equation}
     p_1+p_2 \leq P_{max}
   \label{equ:eq3c2}
   \end{equation}
    \begin{equation}
   b_1+b_2 \leq B
   \label{equ:eq3c3}
   \end{equation}

Since we already know that user 1 is a candidate to the $0$-th rate, there must be a bandwidth-power allocation such that the constraint (\ref{equ:eq3c1}) is satisfied. By solving the optimization problem (\ref{equ:Obj2}-\ref{equ:eq3c3}), we find whether user 2 is a candidate to this rate or not. 

\textbf{Remark }{\it The bandwidth-power allocation for both user 1 and user 2 obtained by solving (\ref{equ:Obj2}-\ref{equ:eq3c3}) overweights the previous decision when considering only user 1. In general, solving the allocation problem at the $u$-th user and $n$-th rate overweights the previous allocation of all users.}

Now, for the $u$-th user and $0$-th rate, we need to solve the following optimization problem:

\begin{equation}
 \underset{{\cal B},{\cal P}}{\textbf{argmax }} b_ulog(1+\frac{\gamma p_u|\alpha_u h_u|_2^2}{N_ob_u}) \label{equ:Obju}
 \end{equation}
 subject to
  \begin{equation}
  b_klog(1+\frac{\gamma p_k|\alpha_kh_k|_2^2}{N_ob_k}) \geq r_0, \forall k < u
    \label{equ:eq4c1}
   \end{equation}
 \begin{equation}
     \sum_{k=1}^{u-1} p_k+p_u \leq P_{max}
   \label{equ:eq4c2}
   \end{equation}
    \begin{equation}
   \sum_{k=1}^{u-1} b_k+b_u \leq B
   \label{equ:eq4c3}
   \end{equation}
If  $b_ulog(1+\frac{\gamma p_u|\alpha_u h_u|_2^2}{N_ob_u})\geq r_0$, then the $u$-th user is a candidate to this rate, otherwise, all users $(u, \dots, U)$ will not be considered for all rates. 

In order to generalize the case to any rate $> r_0$, let's  first define $r_m^k$ to be the maximum rate $< r_n$ for which the user $k$ is a candidate. Then to find out if the $u$-th user is a candidate to the $n$-th rate, we solve the following optimization problem:
\begin{equation}
\underset{{\cal B},{\cal P}}{\textbf{argmax }} b_ulog(1+\frac{\gamma p_u|\alpha_u h_u|_2^2}{N_ob_u}) \label{equ:Objf}
 \end{equation}
 subject to
  \begin{equation}
 b_klog(1+\frac{\gamma p_k|\alpha_k h_k|_2^2}{N_ob_k}) \geq r_m^k, \forall k=1...U, \neq u
    \label{equ:eqfc1}
   \end{equation}
 \begin{equation}
     \sum_{k=1, \neq u}^{U} p_k+p_u \leq P_{max}
   \label{equ:eqfc2}
   \end{equation}
    \begin{equation}
   \sum_{k=1, \neq u}^{U} b_k+b_u \leq B
   \label{equ:eqfc3}
   \end{equation}
If $b_ulog(1+\frac{\gamma p_u|\alpha_u h_u|_2^2}{N_ob_u})\geq r_n$, then the $u$-th user is a candidate to the $n$-th rate, otherwise, it is not a candidate to any rate $\geq r_n$. This means that we exclude all users $(u, \dots, U)$ for all rates $\geq r_n$. Finally, solving the optimization problem (\ref{equ:Objf})-(\ref{equ:eqfc3}) for the highest rate $N$ $(r_N)$ and for the last user that is a candidate to the rate $N-1$ yield the optimal solution to the original problem described in (\ref{equ:mainObj}-\ref{equ:eq1c5}). Next, we proof the optimality of the proposed algorithm in solving the optimization problem described in (\ref{equ:mainObj})-(\ref{equ:eq1c5}).

 \begin{lemma}
Given ($I_{0}^{(u)}, \cdots,I_{n-1}^{(u)}, \forall u \in \{1, \cdots, U\}$), the execution of the algorithm for the $n$-th rate decisions yields the maximum value of $\sum_{u=1}^U I_n^u$. In other words, the proposed algorithm obtains the maximum number of users at rate $n$ as compared to any feasible algorithm which has the same rate decisions of every user up to the rate level $n-1$.
 \label{lem:1} 
\end{lemma}
\textit{Proof}: We first note that using the proposed algorithm, the user $u_s$ is skipped at the $n$-th rate level. i.e, the decision variable of the $n$-th rate and user $u_s$ is equal to $0$ ($I_{n}^{u_{s}}=0$) in two scenarios, which are described as follows.

\noindent {\bf Case 1: } If the user $u_s$ is not a candidate to the $(n-1)$-th rate. i.e, $I_{n-1}^{u_{s}}=0$, no other feasible algorithm can increase the rate of this user, $u_s$, to the $n$-th rate level due to the violation of the constraint (\ref{equ:eq1c2}).

\noindent {\bf Case 2: } There is no feasible bandwidth-power allocation such that constraints (\ref{equ:eqfc1}-\ref{equ:eqfc3}) are satisfied and $b_{u_s}log(1+\frac{\gamma p_{u_s}|h_{u_s}|_2^2}{N_ob_{u_s}}) \geq r_n$

Skips which are due to Case 1 are the same for any feasible algorithm. Thereby, we do not consider skips that are of Case 1.  On the other hand, for the skips of Case 2, we note that the proposed algorithm starts with the user that has the highest channel gain since this user can achieve the $n$-th rate with the lowest resources if that is feasible. Therefore, if there is no bandwidth-power allocation such that this user can achieve the $n$-th rate, there will be no other user can achieve this rate. The algorithm proceeds in descending order with respect to the channel gains. Hence, if user $u_s$ is skipped at the $n$-th rate level, any algorithm would have skipped a user or more with an index $u \leq u_s$. However,  skipping user $u_s$ as compared to user $u \leq u_s$ allows for more bandwidth-power to increase the quality of users with higher channel gains. i.e since user $u_s$ has lower channel gain, he/she needs more resources than user $u \leq u_s$ to reach rate $r_n$. Therefore, skipping the user $u_s$ offers more resources to the remaining users. Thus, we see that the number of $n$-th rate skips in any feasible algorithm can be no less than the proposed algorithm.  $ \small \blacksquare$


\begin{theorem}	
Up to a given rate $N \geq 0$, if $I_{n}^{u^*}$ is the decision variable of the $n$-th rate and $u$-th user ($n \leq N$) that is found by the proposed algorithm, and $I_{n}^{u^\prime}$ is the decision variable of the $n$-th rate and $u$-th user that is found by any other feasible algorithm such that all constraints (\ref{equ:eq1c1}-\ref{equ:eq1c5}) are satisfied, then the following holds when $\lambda$'s satisfy~(\ref{lambdaCons}).
\begin{equation}
\sum_{n=0}^N\lambda_n\sum_{u=1}^U I_n^{u^\prime} \leq \sum_{n=0}^N\lambda_n\sum_{u=1}^U I_n^{u^*} 
\label{thm:thm}
\end{equation}
In other words, the proposed algorithm achieves the optimal solution of the optimization problem~(\ref{equ:mainObj}-\ref{equ:eq1c5}) when $\lambda$'s satisfy (\ref{lambdaCons}).
\label{theorem: theorem11}
\end{theorem}
\textit{Proof}: First, the constraint~(\ref{lambdaCons}) on $\lambda_{0,\cdots, N}$ imposes a strict priority on the rate levels. Therefore, pushing 1 user to the rate $r_n$ yields higher utility than pushing all the other users to rates $> r_n$. Hence, processing rates in their order and sequentially finding the maximum users served at rates $0$ to $N$ in their order will yield the optimal solution to the proposed problem. \textit{Lemma}~\ref{lem:1} proves that the proposed algorithm achieves the maximum number of users at the rate $n \leq N$ given the decisions of all rates up to the rate $n-1$. Projecting that on rate $r_0$, we conclude that the algorithm finds the maximum number of users that can be served at rate $r_0$. Given the $r_0$ decision which is optimal, running the algorithm for $r_1$ will produce the optimal rate decisions for both $r_0$ and $r_1$. Hence, continuing this way up to the rate $r_N$ will yield the optimal solution of all rates and users up to the rate $N$ and that concludes the proof.  

\begin{table}[t]
	\vspace{.05in}
\caption{Proposed Algorithm } 
\begin{tabular}{c l } 
\hline\hline 
  &   \\
 & \textbf{Input}: $r_{0,\cdots,N}$: set of the rates, $\{1,....,U\}$: set of the users,\\ 
 &$B$, $P_{max}$, $H_k\forall k$, $N_0$, $\gamma$ \\ 
& \textbf{Output}: $p_u$ and $b_u$ $\forall u=1, \dots, U$ \\
1 & \textbf{Initialize} : $r_m^u =0 $,  $I_n^u =0 $\\
2 & \textbf{for}  $n=0:N$ \\
3 & \,\,\, $\mathcal{U}=\{1,....,U\}$\\
4 & \,\,\,\textbf{While} ($\mathcal{U}$ in not empty)\\
5 & \,\,\,\,\,\,\, $u\leftarrow$ index of the user with maximum channel gain $\in \mathcal{U}$\\
6 & \,\,\,\,\,\,\,  Solve the optimization problem defined by Eqs(18-21)\\
7  & \,\,\,\,\,\,\, \textbf{if} $(b_u log(1+\frac{\gamma p_u H_u}{N_ob_k}) \geq r_n)$ , then\\
8 &   \,\,\,\,\,\,\,\,\,\, $I_n^u =1, r_m^u=r_n $ \\
9 &   \,\,\,\,\,\,\,\,\,\, $\mathcal{U}=\mathcal{U}-\{u\}$\\
10  &\,\,\,\,\,\,\,\,  \textbf{else}\\   
11 & \,\,\,\,\,\,\,\,\,\,   $I_n^k =0, \forall k \in \mathcal{U}$\\
12 &  \,\,\,\,\,\,\,\,\,\, $\mathcal{U}=\phi$\\
13 & \,\,\,\,\,\,\, \textbf{endif}\\
14   & \,\,\,\,\,\, \textbf{endWhile}\\
15   & \textbf{end}\\
   &  \\ [1ex] 
\hline 
\end{tabular}
\label{table:bbm} 
\end{table}

\section{Numerical Results}
In this section, we provide simulation results for the studied scenario of a single small cell multiuser with video streaming in the downlink. We consider the case of 5 users downloading a video of length 1 minute. The video is divided into 60 chunks, each chunk is of 1 second duration. The video is encoded into 5 rate levels ${[r_1, r_2, r_3, r_4, r_5]}$ which are ${ [0.5, 1, 1.5, 2, 2.5] }$ Mbps, respectively. A user streams with rate 1 ($r_1$) means that it can span a range from 0.5 Mbps to less than 1 Mbps, and it streams with rate 2 ($r_2$), means that it can span a range between 1 Mbps to less than 1.5 Mbps, and so on for the other rate levels. We add rate level 0 ($r_0$) to represent the case when the user cannot achieve the basic rate ($r_1$). The path loss and the Rayleigh fading effects are both considered in the downlink. The path loss gain is computed based on the path loss exponent of 2. The maximum available bandwidth at the BS is 20 MHz and the maximum power is 10 watt. The noise PSD, $N_o$, is taken as $10^{-9}$ for SNR calculation. For the sake of performance comparison with other algorithms, we compare our algorithm with the weighted sum rate maximization when window size, $T$, is 1 and 5. 

The environment of a small cell is considered such that a low mobility scenario can be studied. The channel is considered slowly fading with coherence time that can span a one complete video chunk. Since the proposed algorithm explicitly consider the fairness in the formulation (no weight based on the history is needed), we solve the optimization once per chunk. However, for weighted sum algorithms, in order to ensure fairness among users per chunk we solve the optimization problem 40 time per chunk. In this experiment, the users' distance from the BS is randomly generated  between 5m and 80m and the whole simulation experiment is repeated 100 times.

\begin{figure}[t!b]
 \centering
\includegraphics [trim=.4in .4in .6in .2in, clip, width=.47\textwidth]{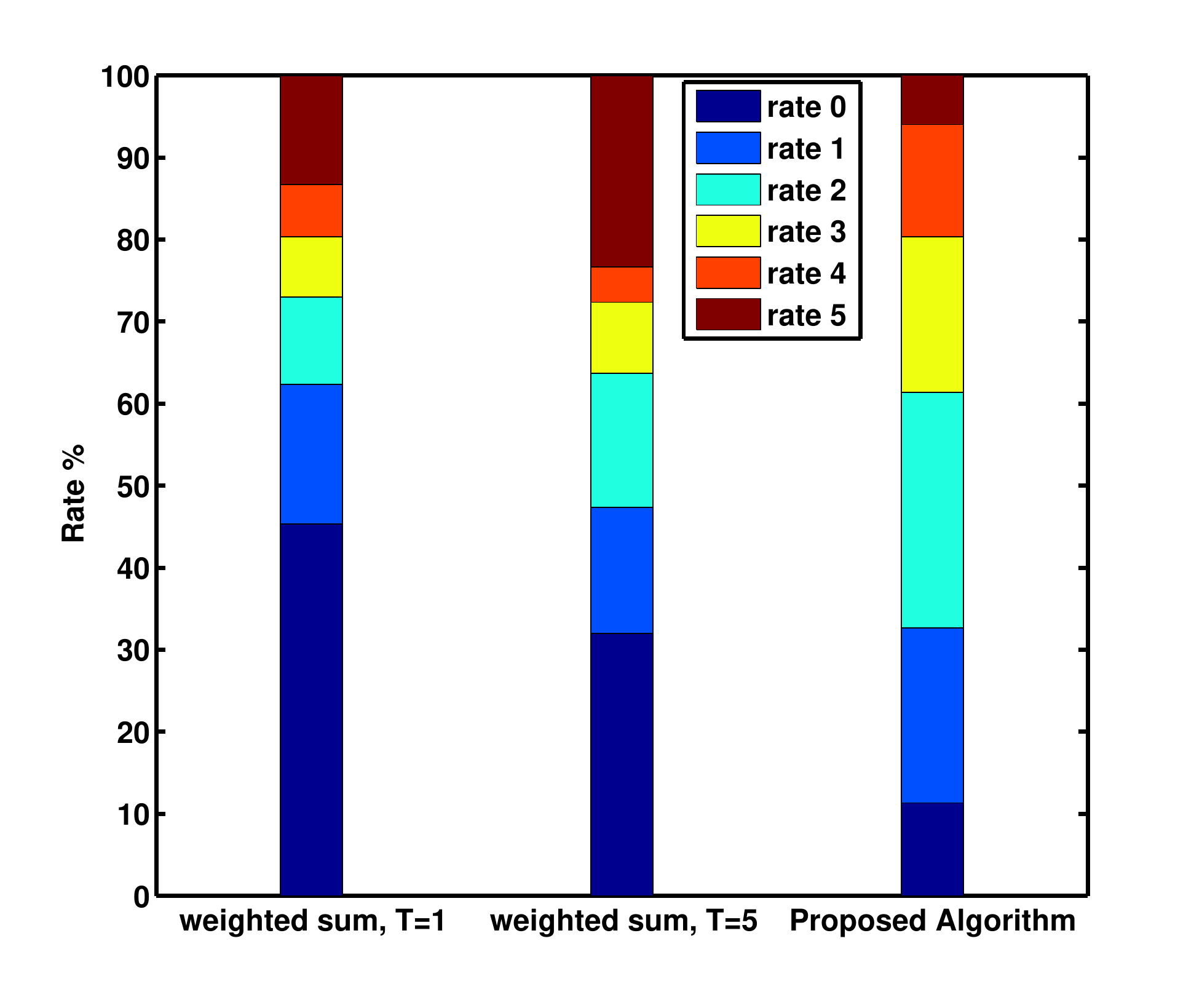}
\caption{ Rate level percentages across all users and all videos rates  }
\label{fig:3}
\end{figure}

Fig. \ref{fig:3} shows the rate percentage across all users and all video rates (${0.5, 1, 1.5, 2, 2.5 }$ Mbps) averaged over all repeated simulations. From this figure, it can be said that users with the proposed algorithm spend less percentage of time experiencing rate 0 level ( i.e. $< 0.5$ Mbps), which is represented by the dark blue color in Fig. \ref{fig:3}, as compared to the weighted sum rate algorithms. This is an advantage of the proposed algorithm from the point of view of the users' QoE. Moreover, the advantage of the proposed algorithm can also be seen in the rate levels 1,2,3, and 4, in which users spend more percentage of time experiencing these rates compared to the other algorithms. The only situation when other algorithms outperform the proposed algorithm is in achieving the rate level 5. However, this advantage comes at the cost of making users experience the rate $r_0$ for more than $30\%$ of the time. The importance of our algorithm is that it reduces the number of time users run below the basic rate, i.e, it minimizes the probability of rebuffing which otherwise significantly degrades the QoE.

\begin{figure*}[!htb]
\setlength{\belowcaptionskip}{-17pt}
 \centering
\includegraphics[trim=1.5in 0.2in 1.5in 0.2in ,clip, width=\textwidth]{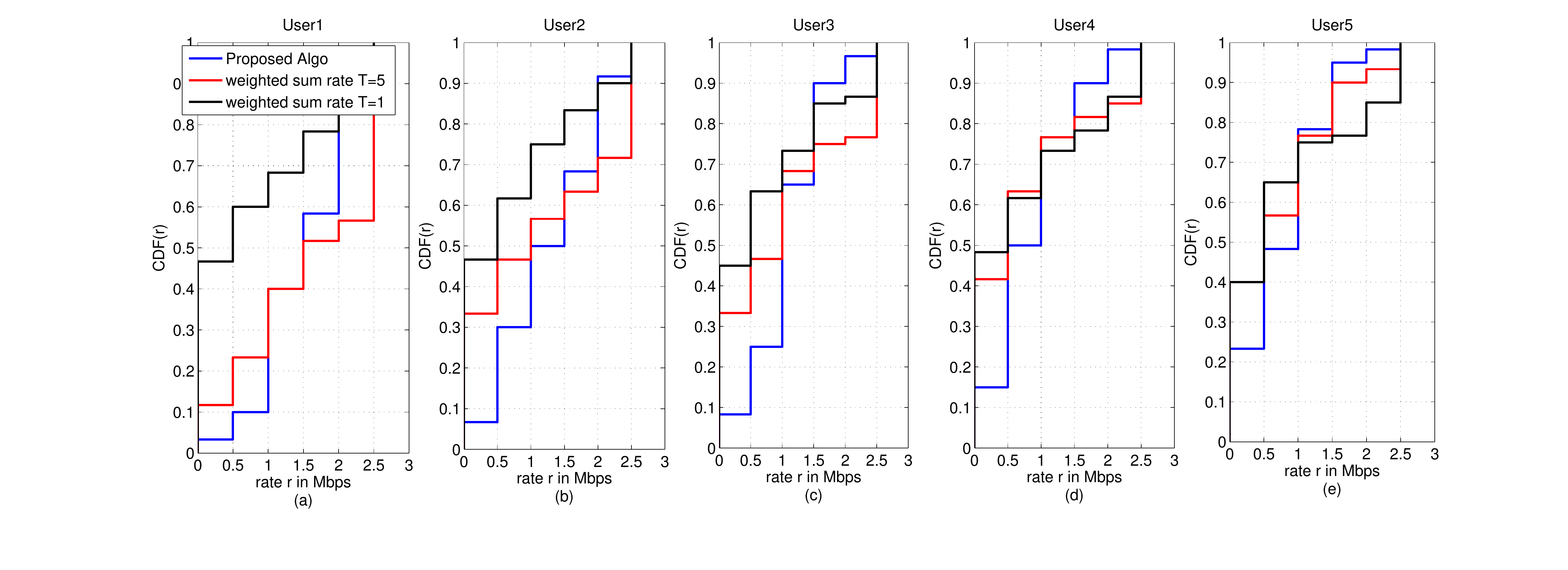}
\caption{ CDF of the achieved rate for all Users}
\label{fig:2}
\end{figure*}

To further analyze the above experiment, we consider a one particular case when the users 1 to 5 are located at distances ${30, 35, 40, 55, 60} $ meters from the BS, respectively.
Fig. \ref{fig:2} shows the CDF of the rate achieved by all users. This figure demonstrates the superiority of the proposed algorithm in guaranteeing a satisfactory rate level for video streaming most of the times, even for the users with the furthest distance from the BS (e.g. users 4 and 5). On the average, our algorithm can guarantee the basic rate or higher for $90\%$ of the time. While on the other hand, users based on the weighted sum rate algorithm suffers to reach the basic rate for about $30 - 40\%$ of the time, on the average. In particular, if we consider users 1 and 5 for all algorithms, we can see from Fig. \ref{fig:2}-a that user 1, which is the closest to the BS, can stream with the basic rate or higher for $97\%$, $89\%$ , and $53\%$ of the time based on the proposed algorithm, weighted sum rate ($T=5$), and weighted sum rate ($T=1$), respectively. Similarly, from Fig (\ref{fig:2})-e, it can be depicted that user 5, which is the furthest user from the BS, can stream with the basic rate or higher for $78\%$, $40\%$ , and $40\%$ of the time based on the proposed algorithm, weighted sum rate ($T=5$), and weighted sum rate ($T=1$), respectively. Similar to the observation we made above in Fig (\ref{fig:3}), the advantage of the weighted sum rate techniques in  Fig's. \ref{fig:2}-c,-d,-e in achieving better rate levels ($>1.5$ Mbps) for users 3, 4, and 5 does not prevent these users from suffering for a noticeable percentage of time. 

\begin{figure}[!htb]
 \centering
\includegraphics [trim=.1in .1in .5in .1in, clip, width=.47\textwidth]{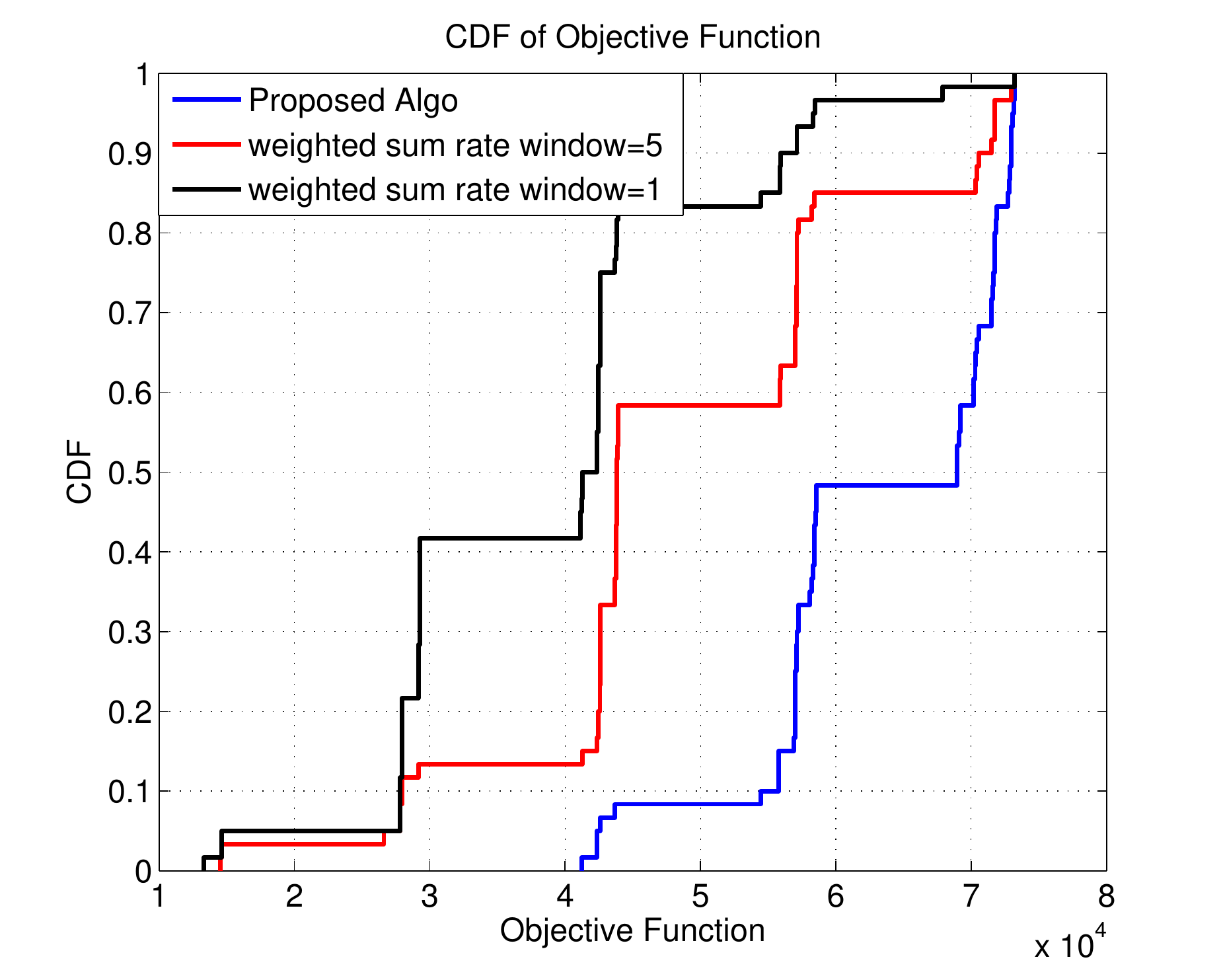}
\caption{ CDF of the objective function value }
\label{fig:4}
\end{figure}

The objective function in \eqref{equ:mainObj} represents the QoE metric in this work. The higher the value of this objective function, the higher the playback rate of each user with guaranteeing proportional fairness among all users. Note that the values of $\lambda_n$ in \eqref{equ:mainObj} are chosen to satisfy \eqref{lambdaCons}, and thus, the values of $\lambda_0 ,\dots,  \lambda_4 $ are chosen, for this simulation experiment, as $[13310, 1210, 110, 10, 1]$, respectively. This choice of  $\lambda_0 ,\dots,  \lambda_4 $ tells that pushing more users to the basic rate level achieves significantly higher objective than pushing one user to the highest rate. Fig. \ref{fig:4} plots the CDF of the objective function for all the considered algorithms. It can be clearly observed that the proposed algorithm significantly outperforms the other algorithms.\\ 

\section{Conclusion}

In this paper, we formulate the joint bandwidth-power allocation for a small cell and video traffic as a non-convex optimization problem whose objective is to optimize the QoE metric that maintains a trade-off between maximizing the playback rate of each user and ensuring users' proportional fairness (PF). We developed a novel algorithm that solves this problem optimally in a polynomial time complexity. Simulation results show that the proposed algorithm outperforms the weighted sum rate in terms of maintaining a high QoE for all users.
\bibliographystyle{IEEEtran}

\bibliography{refs}

\end{document}